\documentclass[9pt,twocolumn,twoside]{osajnl}

\journal{ol} 

\setboolean{shortarticle}{true}

\newcommand{\addtuc}{School of Electrical and Computer Engineering,Technical University of Crete, Chania, Greece 73100}

\newcommand{\addcqt}{Centre for Quantum Technologies, National University of Singapore, 3 Science Drive 2, Singapore 117543}

\newcommand{\addnus}{Department of Physics, National University of Singapore, 2 Science Drive 3, Singapore 117551}

\title{Boosting Topological Zero Modes Using Elastomer Waveguide Arrays}

\author[1,*]{Angelina Frank}
\author[1]{Daniel Leykam}
\author[2,3]{Daria A. Smirnova}
\author[1,4]{Dimitris G. Angelakis}
\author[1,5]{Alexander Ling}

\affil[1]{\addcqt}
\affil[2]{Nonlinear Physics Centre, Research School of Physics, Australian National University, Canberra ACT 2601, Australia}
\affil[3]{Institute of Applied Physics, Russian Academy of Science, Nizhny Novgorod 603950, Russia}
\affil[4]{\addtuc}
\affil[5]{\addnus}

\affil[*]{Corresponding author: angelina.frank@u.nus.edu}

\begin{abstract}
We employ the Su-Schrieffer-Heeger model in elastic polymer waveguide arrays to design and realize travelling topologically protected modes. The observed delocalization of the optical field for superluminal defect velocities agrees well with theoretical descriptions. We apply mechanical strain to modulate the lattices' coupling coefficient. This work demonstrates a novel platform for rapid prototyping of topological photonic devices and establishes strain-tuning as a viable design parameter for topological waveguide arrays.
\end{abstract}

\setboolean{displaycopyright}{true}

\begin{document}

\maketitle

{\bf Introduction.} Elastomeric polymer waveguides present an exciting platform for hybrid photonics. They owe their appeal to a rapid and simple fabrication cycle, mechanical flexibility, as well as the option for direct integration of molecules, nano- and microstructures into the polymer matrix, prompting the development of new schemes for tunable light-matter interaction~\cite{doi:10.1063/1.4998299,Ng:20,Auksztol2019,Frank2022}. These advantages, however, are mitigated by a relatively small refractive index contrast ($\lesssim 0.1$), which places a lower bound of a few micrometers on the waveguide dimensions required for single-mode guiding at visible optical frequencies. It further requires structural elements with millimeter-scale curvatures to avoid bending losses in the guiding structure. Such dimensions can be useful in mesoscopic applications and for practical purposes including optical platform characterization. However, larger dimensions also increase exposure to the environment which can cause higher vulnerability to e.g. fabrication imperfections that raise the noise of acquired data.

Recently, topology has emerged as a promising approach for designing disorder-robust photonic circuits, attracting a variety of experimental demonstrations~\cite{RevModPhys.91.015006}. In planar waveguide arrays light can be robustly localized in mid-gap topological zero modes~\cite{Malkova:09,PhysRevLett.116.163901,Song2020,Guo:20}. A zero mode is a singular localized mode in a photonic system such as a waveguide array that sits in the center of its photonic bandgap. Usually, zero modes have a fixed position, being bound to domain walls formed by structural aperiodicities in the waveguide array, as shown in Fig. 1(a), or the ends of the waveguide array. Adiabatic topological pumping makes use of topological analytical models to engineer the diffraction of light fields and suppress their delocalization. It has therefore attracted interest as a means of robustly manipulating and transporting zero modes~\cite{PhysRevLett.109.106402,PhysRevLett.117.213603,PhysRevB.99.155150,Jurgensen2021,Cheng2022}. In many applications it is desirable to manipulate the topological modes within a finite time or length scale, which has motivated studies of faster protocols based on suppression of non-adiabatic effects~\cite{PhysRevX.3.041017,PhysRevB.91.201404,PhysRevResearch.2.033475}. 

Boosted zero modes that travel laterally at finite speeds have been predicted to preserve their robustness, provided they do not exceed a critical velocity i.e. that the domain wall position changes sufficiently slowly~\cite{PhysRevX.3.041017,PhysRevB.91.201404}. The theory underlying boosted zero modes is based on exploiting the Lorentz invariance of the continuum (long wavelength) description of topological zero modes emerging in condensed matter systems. The continuum description assumes a wide horizontal spread of the light field across an array. By applying a Lorentz boost, travelling zero mode solutions, their velocity-dependent spin and localization, and their robustness to static defects can be studied analytically. Note that here the spin is not an actual photon spin or polarization, but rather refers to the distribution of the light beam between the two sublattices comprising the waveguide array~\cite{RevModPhys.91.015006}. To the best of our knowledge, the velocity-dependent localization and spin of travelling zero mode solutions has not yet been experimentally observed. 

\begin{figure*}
\centering
\includegraphics[width=\linewidth]{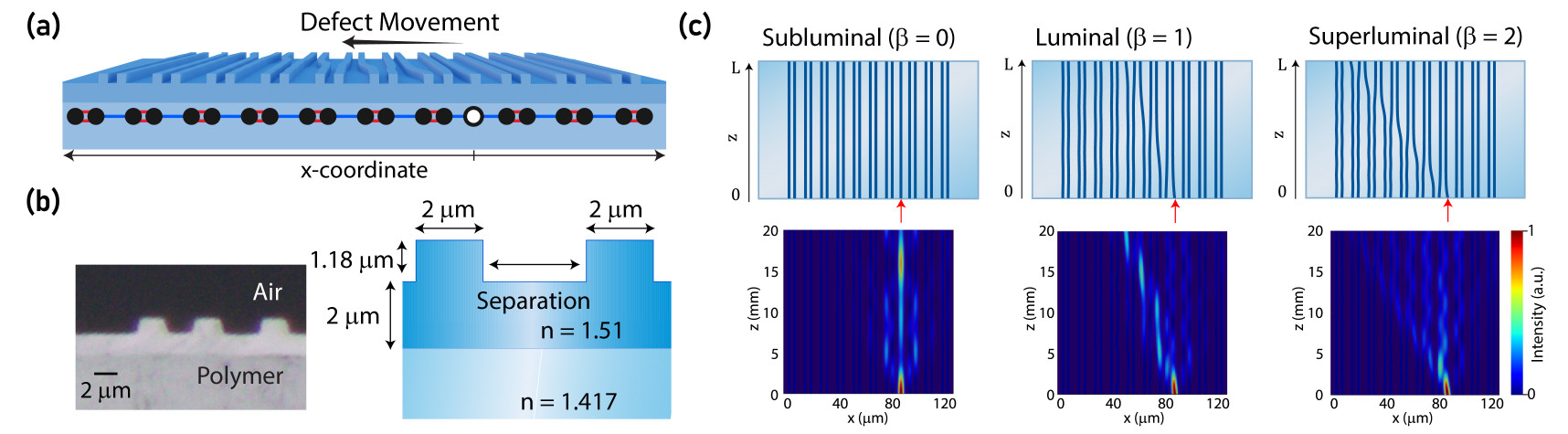}
\caption{(a) Tight binding model (SSH) with modulated nearest-neighbor (NN) couplings, where the white dot indicates the domain wall at the initial injection site. (b) Microscope image of the cross-section and dimensions of the fabricated polymer rib waveguides. The waveguide separation varies between 2 and 5.8 {\textmu}m. (c) 2D Schematic of waveguide arrays with various defect speeds $\beta$ and propagation dynamics simulated using the tight binding model Eqs.~(\ref{eq1},\ref{eq2}).}
\label{fig:fig1}
\end{figure*}

Here, we observe signatures of this behaviour. We use photonic polymer waveguide arrays to experimentally implement boosted topological zero modes. We consider the prototypical Su-Schrieefer-Heeger (SSH) model formed by waveguide arrays with dimerized couplings~\cite{Malkova:09,PhysRevLett.116.163901,Song2020,Guo:20}, creating a topological zero mode by inducing a domain wall in the coupling dimerization. We apply a modulation of the couplings along the longitudinal propagation direction to transport a topological zero mode at finite velocities. The transverse translation of the optical field and its delocalization when the zero mode exceeds a critical velocity (transverse modulation of the coupling) are accurately predicted based on a discrete tight binding model (SSH) describing the coupling between neighboring waveguides and a continuum Jackiw-Rebbi model obtained via a long waveguide expansion~\cite{Angelakis2014,Keil:15,Smirnova2019}. We take advantage of the polymer's flexibility to apply horizontal strain of up to 10$\%$ perpendicular to the direction of propagation and observe a change of the optical field distribution as the effective coupling coefficient decreases. Our experimental results add strain as a potential control parameter for creating tunable topological photonic systems. 

{\bf Model.} We consider the dimerized optical waveguide lattice shown in Fig.~\ref{fig:fig1}(a). Each dimer consists of two  waveguides which are referred to as the two sublattices. This photonic system is experimentally consistent with predictions according to the SSH tight-binding model~\cite{Malkova:09,Ng:20}. When the waveguides are weakly coupled the evolution of the waveguide optical field amplitudes $\psi_n(z) = [a_n(z),b_n(z)]$ through the array, with $z$ being the propagation distance, is governed by the tight binding equations
\begin{align}
    i \partial_z  a_n &= [\kappa - \epsilon_{n-1}(z)/2] b_{n-1} + [\kappa + \epsilon_n(z)/2] b_{n}, \label{eq1} \\
    i \partial_z b_n &= [\kappa - \epsilon_{n+1}(z)/2] a_{n+1} + [\kappa + \epsilon_n(z)/2] a_{n}, \label{eq2}
\end{align}
where $a$ and $b$ denote the optical field amplitudes on the two sublattices, $n$ indexes the unit cells, and $\kappa$ and $\epsilon$ are the average coupling and dimerization strengths, respectively, which are controlled by varying the separation between neighboring waveguides. We normalize the longitudinal and transverse coordinates by the average coupling coefficient and waveguide spacing, respectively, setting $\kappa = 1$ in the following.

We consider the case where the average coupling is kept constant, while an aperiodicity in the dimerization $\epsilon_n(z) = \epsilon_{\infty} \tanh (n - n_0 - \beta z)$ forms a domain wall or defect travelling at a constant defect speed $\beta$. Here, $\epsilon_{\infty}$ is the dimerization strength far from the domain wall, while $n_0$ refers to the initial waveguide number of the defect at $z = 0$. The simulated dynamics of a light beam injected at the domain wall as a function of $\beta$ are presented in Fig.~\ref{fig:fig1}(c). The beam remains localized to the defect only if it travels sufficiently slowly.

The sensitivity of the propagation dynamics to the defect speed can be understood by considering the continuum limit $n \rightarrow x$ of the tight binding equations, which is known as the Jackiw-Rebbi model~\cite{Angelakis2014,Keil:15,Smirnova2019},
\begin{equation} 
[\sigma_z \partial_x + i \sigma_x \partial_z + \sigma_0 \epsilon_{\infty} \tanh (x - x_0 - \beta z)] \psi(x,z) = 0, \label{eq:jr}
\end{equation}
where $x_0$ is the defect position, $\sigma_{z,x}$ are Pauli matrices, and $\sigma_0$ is the identity matrix. When $\beta=0$, ~\eqref{eq:jr} supports modes $\psi(x,z) = \phi(x) e^{-i E z}$ with zero energy ($E=0$) localized to the defect and residing on a single sublattice,
\begin{equation}
    \phi(x) = [ \cosh(x-x_0)^{-\epsilon_{\infty}}, 0].
\end{equation}
This zero mode is robust and exists as long as $\epsilon(x)$ changes sign across the domain wall.

The more general case of a moving domain wall can be analyzed by applying a Lorentz boost~\cite{PhysRevX.3.041017}. Introducing $\gamma = 1/\sqrt{1-\beta^2}$, $x^{\prime} = \gamma(x - \beta z)$, $z^{\prime} = \gamma (z - \beta x)$, and $\psi^{\prime} = S \psi$, where $S = \exp[-\mathrm{arctanh}(\beta)\sigma_y/2]$, \eqref{eq:jr} becomes
\begin{equation}
[\sigma_z \partial_{x^{\prime}} + i \sigma_x \partial_{z^{\prime}} + \epsilon_{\infty} \tanh( x^{\prime}/\gamma - x_0)] \psi^{\prime} = 0, \label{eq:lorentz}
\end{equation}
which takes the same form as \eqref{eq:jr}, but with the domain wall stretched by a factor of $\gamma$ and the sublattice spin rotated by $S$. Thus, as long as the defect speed does not exceed the effective speed of light (subluminal $\beta < 1$) the zero mode persists, but its degree of localization and sublattice spin $\langle \sigma_z \rangle = \int  \phi^*(x)\sigma_z \phi(x) dx = \sqrt{1-\beta^2}$ are both reduced. On the other hand, superluminal defects do not support bound states because one cannot Lorentz boost to a frame in which the domain wall is static. Therefore the input beam as shown in the rightmost lower panel in Fig. 1(c) does not follow the defect, but delocalizes. Further analysis of the continuum Jackiw-Rebbi model can be found in Ref.~\cite{PhysRevX.3.041017}.

{\bf Device design and fabrication.} To experimentally study the dynamics of the boosted zero modes we use an array of rib waveguides with spacings between 5.8 {\textmu}m and 2 {\textmu}m. Figure ~\ref{fig:fig1}(b) shows a cross section and schematic of the waveguides. The waveguides have a width of 2 {\textmu}m, and a rib height of 1.18 {\textmu}m. The guiding layer's slab height is set to 2 {\textmu}m, which is engineered to maintain single-mode guiding while ensuring appreciable mode coupling between adjacent waveguides. The array spans a longitudinal propagation length of $\approx$ 2 cm. The initial mold for the polymer was written via direct photolithography. A heat curing polymer (GelestOE50, Mitsubishi Chemicals, refractive index = 1.51) was spin-coated to a thickness of 2 {\textmu}m onto this substrate and cured for 4 hrs at 55 $^{\circ}$C. A second polymer layer (Sylgard 184, Dow, refractive index = 1.42) was applied on top and cured for 1 hr at 70 $^{\circ}$C. The structure was immersed in dimethyl sulfoxide (DMSO) overnight for liftoff. In contrast to semiconductor platforms such as silicon nitride with a turnaround time of up to several months, the entire fabrication cycle here does not exceed several days. 

We employed full-wave simulations (Lumerical) to obtain the dependence of the waveguides' coupling coefficients as a function of their separation. For evanescent coupling, the coupling strength depends on their centre-to-centre separation $d$ according to
\begin{equation}
C(d) = C_0 \exp(-\alpha d), \label{eq:coupling}
\end{equation} 
where $C_0 \approx 1.6$ mm$^{-1}$ and $\alpha \approx 0.3$ {\textmu}m$^{-1}$. The minimum centre-to-centre separation of $4$ {\textmu}m allows for coupling strengths of up to $\approx 0.45$ mm$^{-1}$; for our designs we set $\kappa = \epsilon_{\infty} = 0.3$ mm$^{-1}$ and fabricate arrays of 21 waveguides with different defect speeds $\beta$; for these parameters $\beta = 1$ corresponds to a defect ``speed'' of 1.8 {\textmu}m / mm in physical units.

\begin{figure}
\centering\includegraphics[width=\linewidth]{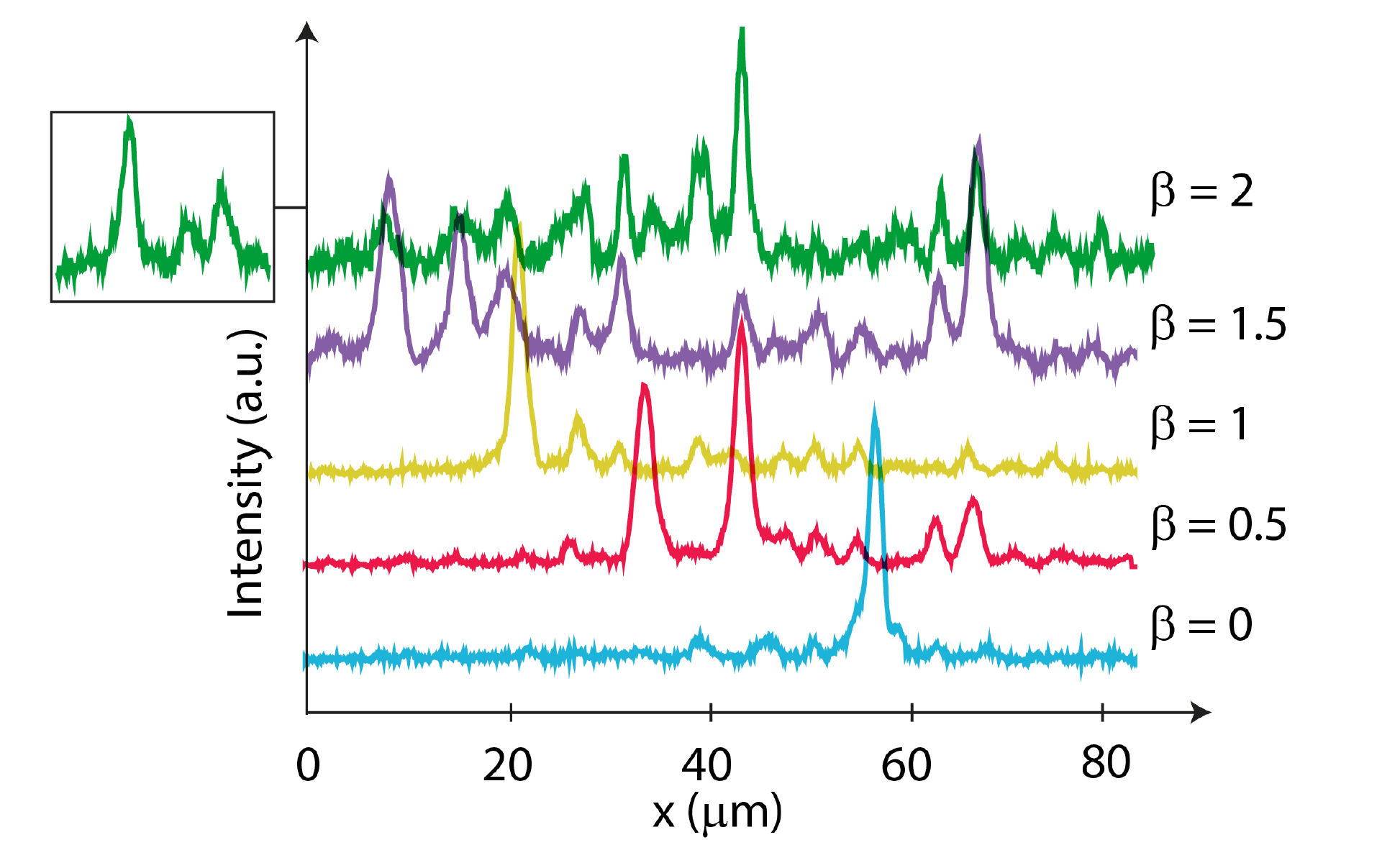}
\caption{Average output intensity profiles for various defect speeds $\beta$. The defect in the array shifts towards the left with increasing $\beta$. Delocalization is apparent for $\beta > 1$. The inset for $\beta$ = 2 shows the output at the defect location after propagation. For $0 \leq \beta \leq 1.5$ and $\beta =2$ the traces show the average of four and two different array sets, respectively.}
\label{fig:fig2}
\end{figure}

{\bf Delocalization of boosted defect modes.} To characterize the experimental signature of varying defect speeds we edge-couple a TM-polarized 633 nm light source to the the defect waveguide at $z=0$ and measure the output profile, i.e. the lateral intensity distribution after propagation of the input beam through the structure, using a microscope objective and CCD camera. As visible in Fig.~\ref{fig:fig2}, the light field is strongly localized to the stationary defect, undergoes a horizontal translation following the defect for $0 < \beta < 1$, and delocalizes for $\beta > 1$.

To quantitatively characterize the output intensity distributions, we apply Gaussian fits to the output peaks to obtain intensities at the individual waveguides, $|a_n|^2$ and $|b_n|^2$, which we normalize such that $\sum_n (|a_n|^2 + |b_n|^2) = 1$. We then characterize the distributions using three measures: (1) The inverse participation ratio (IPR),
\begin{equation}
\mathrm{IPR} = 1/\sum_n (|a_n|^4 + |b_n|^4), \label{eq:IPR}
\end{equation}
which measures the number of waveguides with appreciable intensity; a larger value corresponds to a wider distribution i.e. greater spread of the optical field; (2) The beam centre of mass
\begin{equation}
\langle x \rangle = \sum_n [ n|a_n|^2 + (n+\frac{1}{2})|b_n|^2 ], 
\end{equation}
and the sublattice spin
\begin{equation}
\langle \sigma_z \rangle = \sum_n (|a_n|^2 - |b_n|^2),
\end{equation}
which measures the imbalance between the total intensity on each sublattice. 

\begin{figure}
\centering
\includegraphics[width=\linewidth]{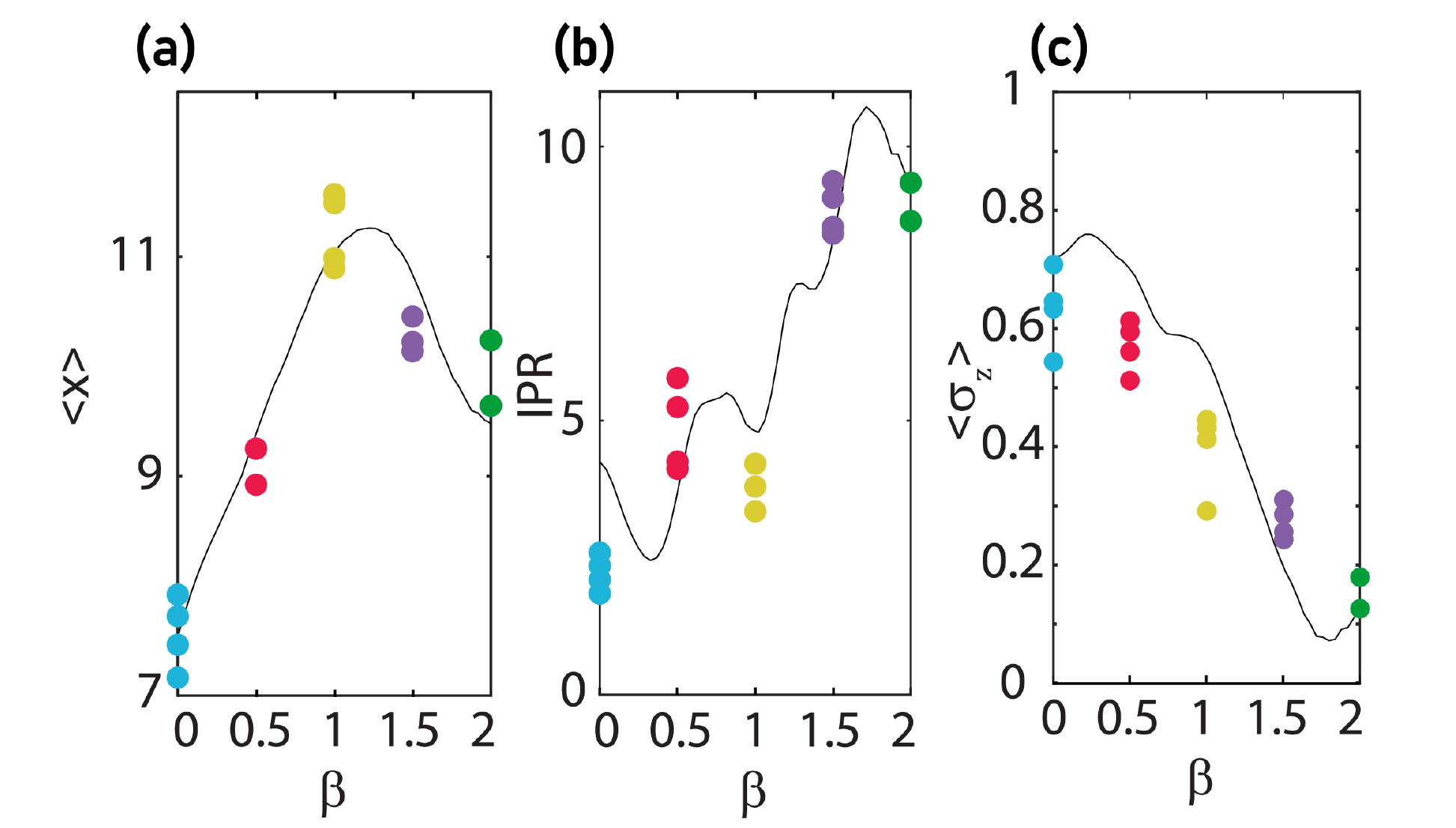}
\caption{Comparison between output beam profile measures obtained from the experimental observations (points) and tight binding model simulations (curves). (a) Centre of mass $\langle x \rangle$. (b) Inverse participation ratio IPR. (c) Sublattice intensity imbalance $\langle \sigma_z \rangle$, expressing the total intensity difference between the even and odd waveguides.}
\label{fig:fig3}
\end{figure}

We compare measured waveguide intensities with profiles computed numerically using the tight binding model. To reproduce the experimental observations, we found that it is essential to include negative next-nearest neighbor coupling terms, whose strength $\approx 0.046$ mm$^{-1}$ also follows \eqref{eq:IPR}. This breaks the chiral symmetry responsible for the topological protection, slightly shifting the defect mode away from zero energy.

Figure ~\ref{fig:fig3} compares the measured and numerically-simulated output measures as a function of the defect speed. Changes in the beam centre of mass away from the input waveguide are reflected in the output beam displacement. Notably, the maximal output beam shift is obtained for luminal defect speeds $(\beta \approx 1)$, in good agreement with the tight binding model simulations. For $\beta > 1$ the beam is no longer able to keep up with the defect and instead undergoes a sharp delocalization transition. Finally, the output sublattice imbalance $\langle \sigma_z \rangle$ steadily decreases as a function of the defect speed, which is consistent with the spin rotation arising from the Lorentz boost in ~\eqref{eq:lorentz}.

\begin{figure*}
\centering
\includegraphics[width=0.8\linewidth]{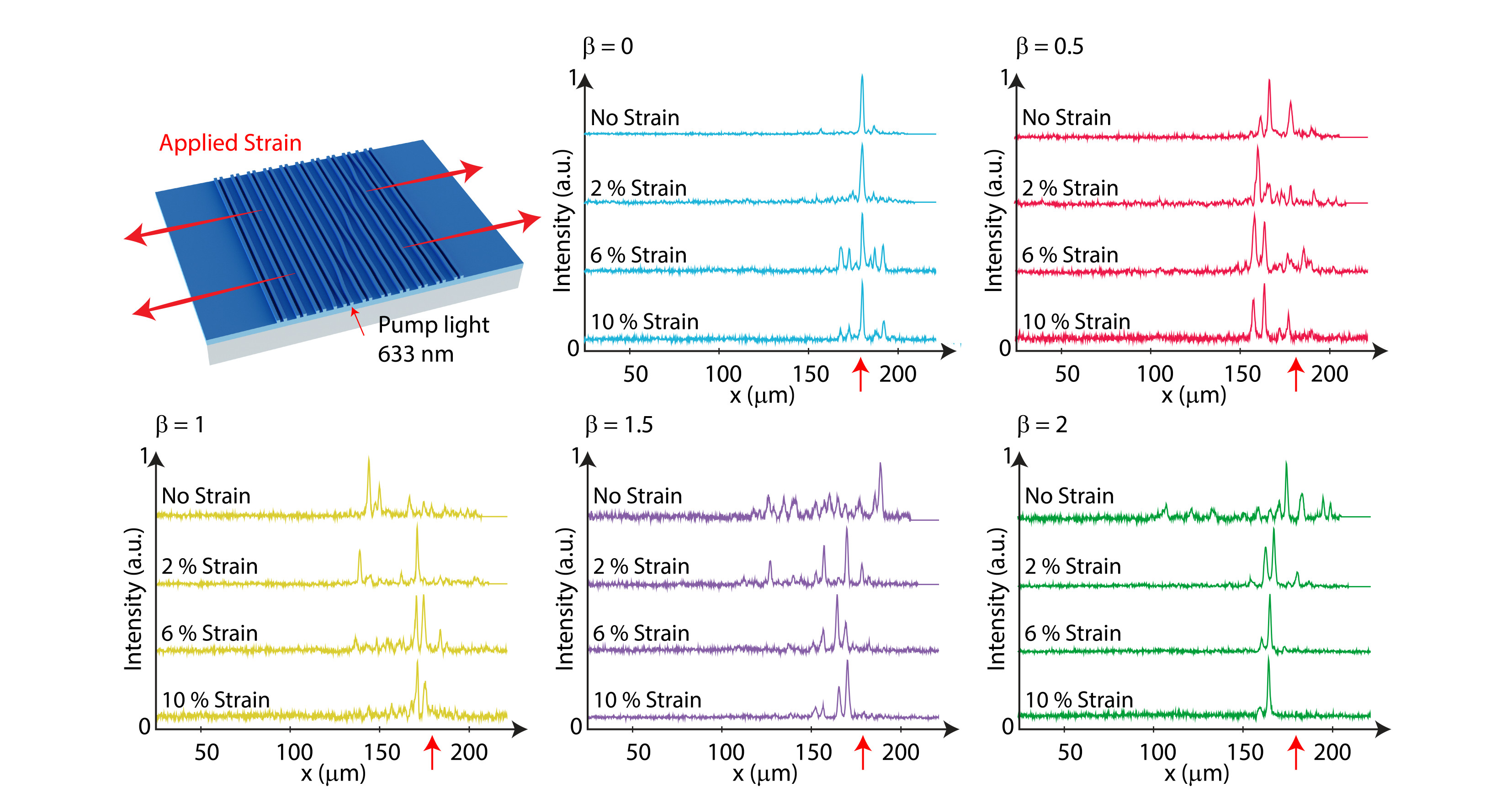}
\caption{Schematic of the strain applied to the sample and measured output intensity profiles for various $\beta$ and levels of applied strain. The strain is expressed as the relative elongation of the chip with respect to its original width. A strain of 100$\%$ would indicate an elongation of two times the original width. Red arrows mark positions of the input beam.}
\label{fig:fig4}
\end{figure*}

{\bf Strain tuning.} It was previously shown that applying horizontal strain perpendicular to the elastomeric waveguide arrays can be used to tune the coupling coefficient~\cite{Ng:20}. For evanescent coupling of the form ~\eqref{eq:coupling} the effect of strain $s$ can be modelled as $C(d,s) = C_0 \exp(-\alpha d(1+s))$, such that $\partial_s C = - \alpha d C$; waveguides that are further apart experience a larger relative reduction in their coupling. Thus, the strain will not only reduce the average coupling $\kappa$, but also increase the relative strength of the dimerization $\epsilon / \kappa$, reducing the bulk mode dispersion and resulting in stronger localization of the output beam to the input waveguide. In other words, the strain has the effect of reducing the system to decoupled waveguide dimers (in the bulk) and a trimer (in the vicinity of the defect). This can be seen in the experimentally-measured output profiles in Fig.~\ref{fig:fig4}: As the strain is increased we observe that the output beam localizes in the vicinity of the initially excited site. For higher defect speeds this effect is more pronounced due to different initial separations at the injection site. Thus, using strain we can tune between collective propagation dynamics involving the entire array and the discrete non-topological dynamics of a few coupled waveguides. This feature offers potential for strain-tunable topological protection. We apply strain by clamping the chips along both edges on a mechanical translation stage with two anchor points that are then being increasingly separated.

{\bf Conclusion.} We have observed experimental signatures of boosted topological zero modes exhibiting velocity-dependent (de-)localization in a strain-tunable, photonic elastomer platform. To the best of our knowledge, this is the first time that such travelling modes were observed in a photonic system. Introducing topological photonics to polymer waveguide systems may be beneficial from two angles: Firstly, topological protection can enhance the platform's robustness and simplifies the design process. Secondly, the study of topological concepts can be facilitated through a photonic platform that not only allows for fast prototyping, but also enables the exploration of topological effects in hybrid photonic systems. In particular, integration of 2D materials and molecular dopants into the polymer matrix may provide a novel platform for the study of quantum and nonlinear topological wave effects~\cite{Smirnova2020}.

\begin{backmatter}
\bmsection{Funding} 
We acknowledge support from the National Research Foundation, Prime Minister’s Office, Singapore, under the Research Centres of Excellence and Medium-Sized Centre programmes, the Ministry of Education, Singapore, under AcRF Tier 3 (MOE2018-T3-1-005), and the Australian Research Council (grant DE190100430)

\bmsection{Acknowledgments} The authors are very thankful to Prof. Eda Goki, Justin Zhou and Jian Linke for facilitating the fabrication processes and to James A. Grieve for many fruitful discussions.

\bmsection{Disclosures} The authors declare no conflicts of interest.

\bmsection{Data availability} Data underlying the results presented in this paper are not publicly available at this time but may be obtained from the authors upon reasonable request.

\end{backmatter}

\bibliography{sample}

\bibliographyfullrefs{sample}

\end{document}